\documentclass[letterpaper,superscriptaddress,aps,twocolumn]{revtex4}
\usepackage{graphicx,color,upgreek}
\usepackage{mathtools}
\usepackage{amsmath}
\usepackage{layouts}
\usepackage{float}
\usepackage{hyperref}
\hypersetup{
    colorlinks=true,
    linkcolor=blue,
    filecolor=magenta,      
    urlcolor=cyan,
    pdftitle={Overleaf Example},
    pdfpagemode=FullScreen,
    }
\usepackage{makecell}
\usepackage[abs]{overpic}
\usepackage{helvet}
\usepackage{graphicx,color,natbib}
\usepackage{placeins}
\usepackage[dvipsnames]{xcolor}
\usepackage{array, booktabs, makecell}
\usepackage{siunitx}
\usepackage{longtable}
\usepackage{tabularx}
\usepackage{booktabs}
\let\oldAA\AA
\renewcommand{\AA}{\text{\normalfont\oldAA}}
\usepackage{textcomp}
\usepackage{color}
\usepackage{amssymb,amsmath} 
\usepackage{siunitx}

\begin{document}

\title{Polymorph Selection in Charged Colloids in the Second Nucleation Step}

\author{C. Patrick Royall}
\affiliation{Gulliver UMR CNRS 7083, ESPCI Paris, Universit\' e PSL, 75005 Paris, France.}
\email{paddy.royall@espci.psl.eu}

\begin{abstract}
We study polymorph selection in a model of charged colloids, with a focus on the higher-order structure prior to and during nucleation. Specifically, we carry out molecular dynamics simulations of a repulsive Yukawa system with a slightly softened (Weeks-Chandler-Andersen) core. We consider the case where the interaction is long-ranged and the BCC crystal is stable, and also intermediate- and short-ranged cases where the FCC crystal is stable.  We use two methods for structure identification, the topological cluster classification (TCC) [A. Malins \emph{et al., J. Chem. Phys.} \textbf{139}, 234506 (2013)] and the bond orientational order parameter analysis of W. Lechner and C. Dellago [\emph{J. Chem. Phys.} \textbf{129}, 114707 (2008)]. Under conditions of high supersaturation, appropriate to experiments with colloids, we find that the system forms a precursor state in which the particles are hexagonally ordered. ~That is to say, the precursors are indistinguishable from an HCP crystal using the  bond orientational order parameters.  This ordering occurs at state points both when the body-centred cubic crystal is the stable phase, and also when the face-centred cubic crystal is stable. In all cases, the stable polymorph forms from the precursor phase in a second stage. Although at freezing, the fluid is very much more ordered when the interactions are short-ranged (when FCC is stable), at the supersaturations where nucleation occurs in our simulations, the higher-order structure of the metastable fluids is almost identical for the long-, short-, and intermediate-ranged systems when measured with the TCC.
\end{abstract}

\maketitle

\section{Introduction}
\label{sectionIntroduction}

Crystallisation is an everyday phenomenon and although it has been studied for many years, it remains a challenging problem. Crystals may form via homogeneous nucleation, which is a rare event that occurs on a microscopic lengthscale and timescale~\cite{sear2012}. It is hard to access such phenomena in conventional materials, but experiments with colloidal dispersions in which the individual particles can be tracked~\cite{gasser2001,taffs2013,tan2014,ivlev,royall2024} and computer simulation~\cite{sosso2016} play an important role in understanding crystal nucleation.

Almost all materials have more than one crystal polymorph with differing thermodynamic stability, depending on the state point. Crucially, the thermodynamic stability may not predict the kinetic stability of the material~\cite{sear2012}. This can have spectacular consequences, for example in the well--known case of the anti--AIDS drug Ritonavir which formed a previously unknown polymorph, rendering it unusable with very significant patient care consequences, not to mention the cost, of \$250M~\cite{bucar2015,bauer2001,mithen2015}. Other examples include biomineralisation in which the less stable polymorph of calcium carbonate, aragonite, is formed in a biological setting, yet in the laboratory the stable form calcite is usually found, at least at temperatures appropriate to corals and shells~\cite{meldrum2008,zeng2017}.

\begin{figure}
\includegraphics[width=85 mm]{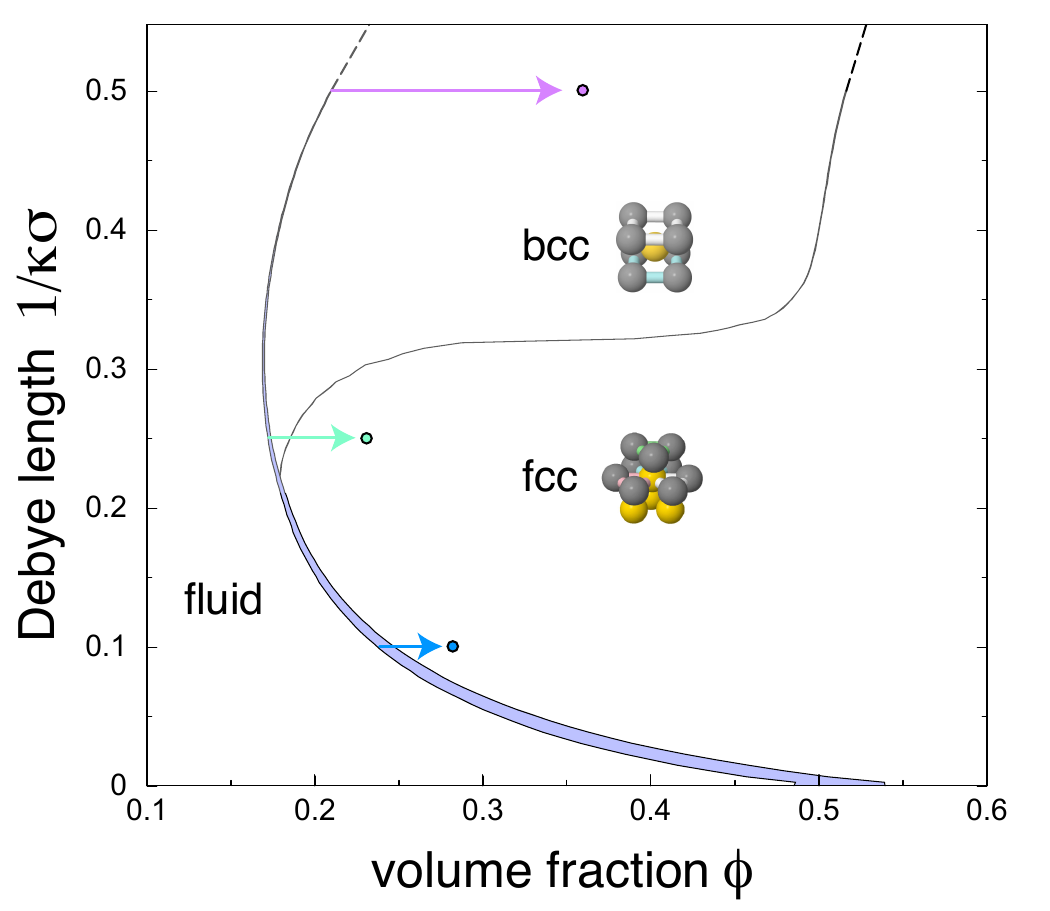}
\caption{Phase diagram of the hard core-Yukawa system for a contact potential $\beta \epsilon_\mathrm{yuk} =39$. Reproduced with permission from A.-P. Hynninen and M. Dijkstra, \emph{Phys. Rev. E.}  \textbf{68}, 021407 (2003). Copyright 2003 American Physical Society. State points shown are the weakest supercooling that crystallised in this work. Arrows indicate the degree of supercooling with respect to the phase boundary. Violet data is the long-ranged case with BCC the stable polymorph. Blue data is the short-ranged case with BCC the stable polymorph.  Green data is the intermediate-ranged case where, for the supersaturations studied here, FCC is the stable polymorph. FCC.}
\label{figPhase}
\end{figure}

Understanding of such complex phenomena can be aided by the use of model systems. Here we consider the Yukawa model with a softened core, which has two polymorphs. This model is well approximated by charged colloids, provided that the degree of electrostatic charge is not too high so that the linear Poisson-Boltzmann approximation in the Derjagiun-Landau-Vervey-Overbeek theory can be used~\cite{royall2003,royall2006,riosdeanda2015}. Due to the finite size of the colloidal particles, a hard core is often included, which significantly changes the phase behaviour when the Yukawa contribution is weak or short-ranged~\cite{royall2006,taffs2013}. Often, (when the Yukawa interactions are strong enough and long-ranged enough that the cores do not come into contact) the hard core Yukawa model gives results that are indistinguishable from those in the absence of a 
core~\cite{hynninen2003}. As shown in the phase diagram in Fig.~\ref{figPhase}, these polymorphs are FCC in the case that the screening is strong (short range) and BCC when the screening is weak (long range). The triple point (fluid-FCC-BCC coexistence) has also been accessed in experiments, where a rather low surface tension between the polymorphs was found~\cite{chaudhuri2017}.

Now ~\citet{alexander1978} argued in 
general mean-field terms that for simple liquids with a spherically symmetric interaction, BCC is typically to be expected as the first polymorph that forms, and the repulsive Yukawa system forms a suitable testbed for such a prediction. Polymorph selection in Yukawa systems has been studied using computer simulation by~\citet{desgranges2007jcp}, who found that when the interaction was long-ranged, BCC nucleated, yet when the interaction was short-ranged, in addition to the stable FCC phase, BCC was also found, consistent with the ideas of~\citet{alexander1978}. BCC was also found in the FCC-stable regime by ~\citet{krazter2015} and also in experiment~\cite{xu2010}.

In some model systems, precursors, i.e. ordered regions with structure distinct from that of the nucleus have been found \emph{before} the nucleus forms. In particular~\citet{russo2012scirep} found a hexagonal ordering ahead of FCC nucleation in hard spheres, ~\citet{lechner2011} found hexagonal ordering in the Gaussian core model in the case that the stable phase was both FCC and BCC. In the same model,~\citet{russo2012sm} additionally found a BCC precursor when the stable phase was FCC. In an experimental \emph{tour-de-force}, Tan \emph{et al.}~\cite{tan2014} showed evidence for a precursor in experiments on charged colloids, ie a hard core Yukawa system. ~\citet{russo2016} also found a precursor in simulations of water. More recently, using a range of order parameters de Jager \emph{et al.}~\cite{dejager2023}, found no evidence for precursors in a hard core Yukawa system. Mithen \emph{et al.}~\cite{mithen2015} performed large-scale simulations of the Gaussian Core Model, and found (unlike refs.~\cite{lechner2011,russo2012sm}) that nuclei of mixed BCC, FCC and HCP composition formed. Clearly, the picture that emerges from this selection of model systems is not entirely consistent. Variations in state point and perhaps order parameter used may play a role, a topic to which we return at the end of this work.

\begin{figure}
\includegraphics[width=85 mm]{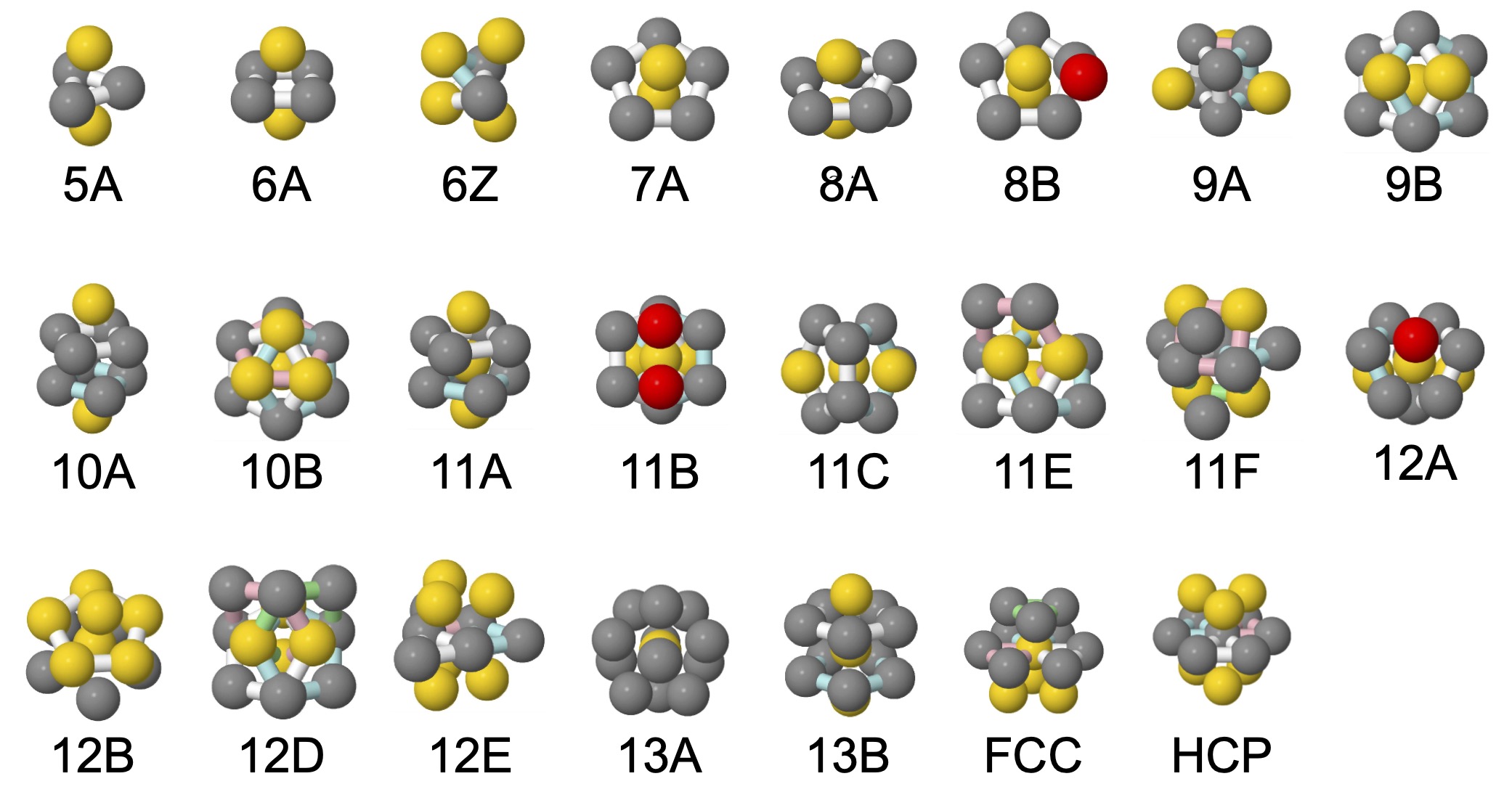}
\caption{Higher-order structures identified by the topological cluster classification and crystals.
Numbers correspond to the number of particles in the cluster. Letters to the range of the Morse potential $\rho_0$ for which these are minimum energy clusters (See Eq.~\ref{eqMorse} in the Appendix)~\cite{doye1995}. Reproduced from [A. Malins \emph{et al., J. Chem. Phys.} \textbf{139}, 234506 (2013)], with the permission of AIP Publishing.}
\label{figTCCStructures}
\end{figure}

Due to their mesoscopic size, colloidal particles exhibit dynamics that are very much slower than in molecular systems~\cite{russel}. In fact, even without rare-event sampling methods often used to study molecular systems, it is possible to directly compare nucleation in colloids with brute force (unbiased) computer simulations~\cite{taffs2013}, and this is the approach we shall employ. Here we make a detailed study of the structure of nuclei and precursors in a Yukawa system using molecular dynamics simulations. simulations, 
We consider cases where both the BCC and FCC form the stable polymorph, with a long-ranged and short-ranged Yukawa interaction respectively. We also investigate an intermediate case close to the BCC-FCC phase boundary. We use two order parameters to probe the higher-order structure of the nucleating system. Firstly, we investigate the structure of the supersaturated fluid using the topological cluster classification (TCC), which identifies geometric motifs whose bond network is identical to minimum energy clusters of the variable-ranged Morse potential (Fig.~\ref{figTCCStructures})~\cite{malins2013tcc}. Secondly, we use the bond orientational order (BOOP) parameter method~\cite{steinhardt1983}. In particular we implement the variant introduced by~\citet{lechner2008} in which second-nearest neighbours are also considered. In the case for both BCC and FCC stable state points, the formation of the nucleus is preceded by a hexagonally ordered structure. This then gives way to the stable polymorph in a second nucleation step.

This work is organised as follows. In the methodology section (Sec. ~\ref{sectionMethods}), we first outline  the molecular dynamics simulations used (Sec. ~\ref{sectionMolecular}) before describing briefly the topological cluster classification  (Sec.~\ref{sectionTopological}) and then the bond orientational order parameter used (Sec.~\ref{sectionBOOP}). The results are presented in Sec.~\ref{sectionResults}. This is broken up into the results from the TCC analysis of the higher-order structure of the bulk fluid prior to nucleation in Sec.~\ref{sectionHigher}, followed by the study of the nucleation process using the BOOP (Sec.~\ref{sectionBOOP}). The structure of the precursors and the surrounding fluid is considered in Sec.~\ref{sectionTCCprecursor}. Finally we discuss our findings in Sec.~\ref{sectionDiscussion} and conclude in Sec.~\ref{sectionConclusion}.

\section{Methods}
\label{sectionMethods}

\subsection{Molecular dynamics simulation}
\label{sectionMolecular}

As our model, we use a Yukawa system with a slightly softened core, which is representative of charged colloids~\cite{taffs2013,royall2013myth}.

\begin{equation}
\label{eqHYuk}
\beta u_\mathrm{hyuk}(r) = \beta u_\mathrm{wca}(r) +  \beta u_\mathrm{yuk}(r) 
\end{equation}

\noindent
where $\beta$ is the inverse of the thermal energy $k_BT$.

For the core, we take the Weeks-Chandler-Andersen truncation of the Lennard-Jones interaction~\cite{weeks1971}.

\begin{align}
\label{eqWCA}
u_\mathrm{wca}(r) &=
\begin{cases}
4 \varepsilon_\mathrm{wca}[(\frac{\sigma'}{r})^{12} - (\frac{\sigma'}{r})^6] + \varepsilon_\mathrm{wca}  & \ r \le 2^{\frac{1}{6}}\sigma' \\[2px]
0 &\ r > 2^{\frac{1}{6}}\sigma'
\end{cases}
\end{align}

\noindent
where  $\varepsilon_\mathrm{wca}=10 k_BT$ is the interaction energy and $r$ is the separation between two particles of diameter $\sigma'$.

The Yukawa potential reads

\begin{equation}
\label{eqYuk}
\beta u_\mathrm{yuk}(r) = \beta \varepsilon_\mathrm{yuk} \frac{\exp\left(-\kappa(r-\sigma') \right)}{r/\sigma'} 
\end{equation}

\noindent
where $\kappa$ is the inverse of the Debye screening length. $\beta \varepsilon_\mathrm{yuk}$ is the potential at contact.

To express the potential at contact, we set the unit of length to be the effective hard sphere diameter, defined according to Barker and Henderson~\cite{barker1976}.

\begin{equation}
\label{eqSigmaEff}
\sigma_\mathrm{eff} = \int_0^{\infty} 1 - \exp \left(-\beta u_\mathrm{wca}(r) \right) dr
\end{equation}

\noindent
Here $\sigma_\mathrm{eff}\approx1.0786$.

We use $\sigma_\mathrm{eff}$ as our unit of length throughout. Henceforth, we drop the subscript $x_\mathrm{eff}$ and define the interaction parameters as $\kappa\sigma =2.0,4.0$ and $10.0$ for long, intermediate, and short-ranged cases respectively.short ranged case. Throughout, we set the contact potential $\beta \varepsilon_\mathrm{yuk} = 39.0$. The phase diagram for this model was determined by Hynninen and Dijkstra~\cite{hynninen2003} and is shown in Fig.~\ref{figPhase} where $\phi=\pi\rho/(6\sigma^3)$ is the volume fraction and $\rho$ is the number density. Hynninen and Dijkstra~\cite{hynninen2003} consider an perfectly hard core. However for our parameters, we see no reason to suppose that our system would behave in a significantly different manner.

We use molecular dynamics simulations. While colloids of course exhibit overdamped dynamics, in dense fluids dynamical quantities have been shown to exhibit remarkably little dependence on the particular dynamics used~\cite{berthier2007jpcm} and indeed molecular dynamics simulations have been successful directly compared with colloid experiment~\cite{royall2007jcp}. For the nucleation simulations presented here, we used the NPT ensemble. We set the system size $N= 11664$ or $65536$ for the long-ranged case $\kappa\sigma = 2.0$, $N= 10976$ or $87808$ for the intermediate-ranged case $\kappa\sigma = 4.0$, and $16384$ or $87808$ for the short-ranged $\kappa\sigma = 10.0$ case. These system sizes are ``magic numbers'' in the sense that BCC and FCC crystals can form perfect crystals for the long-ranged ($N=11664, 65536$) or intermediate- and short-ranged ($N=10976, 16384, 87808$) cases respectively. We run the system for up to $3\times 10^6$ Lennard Jones time units, which we take as our unit of time throughout. No qualitative difference was observed between the different system sizes. Renderings are presented for $N=10976, 11664, 16384$ data.

The simulations were carried out using the LAMMPS package~\cite{plimpton1995} and we performed at least 6 runs for each state point. The system was prepared as a set of random coordinates, minimised under Eq.~\ref{eqWCA} to remove overlaps between particles before the runs were started. We determined the equation of state for the supersaturated fluid, from which we selected the volume fraction at which to launch the nucleation runs.

We find that freezing occurs on the simulation timescale for volume fractions $\phi=0.356...0.389$ and $0.279...0.300$ for the long-ranged ($\kappa\sigma = 2.0$), $0.230..0.284$ for the intermediate ($\kappa\sigma = 4.0$) and $\phi=0.356...0.389$ for the short-ranged ($\kappa\sigma = 10.0$) systems respectively. At higher volume fractions, crystallisation occurs rapidly, in a ``spinodal-like'' manner, as is the case for hard spheres~\cite{zaccarelli2009,royall2024}.

\subsection{Topological cluster classification}
\label{sectionTopological}

\begin{figure*}
\includegraphics[width=170 mm]{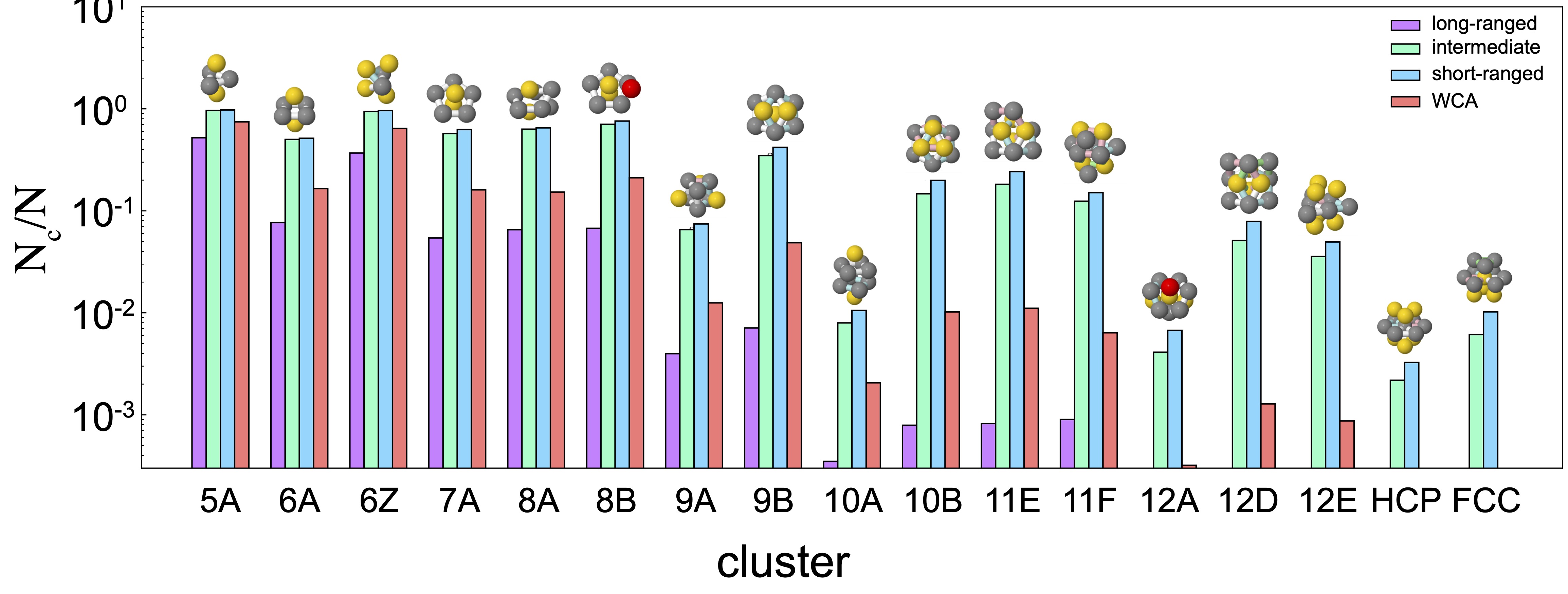}
\caption{Population of particles in clusters identified by the topological cluster classification at freezing. Shown are data  for the long-ranged system (violet), intermediate (green), 
short-ranged (blue) and WCA (red). The WCA system is mapped to the freezing point of hard spheres.}
\label{figTCCFreezing}
\end{figure*}

\begin{figure*}
\includegraphics[width=170 mm]{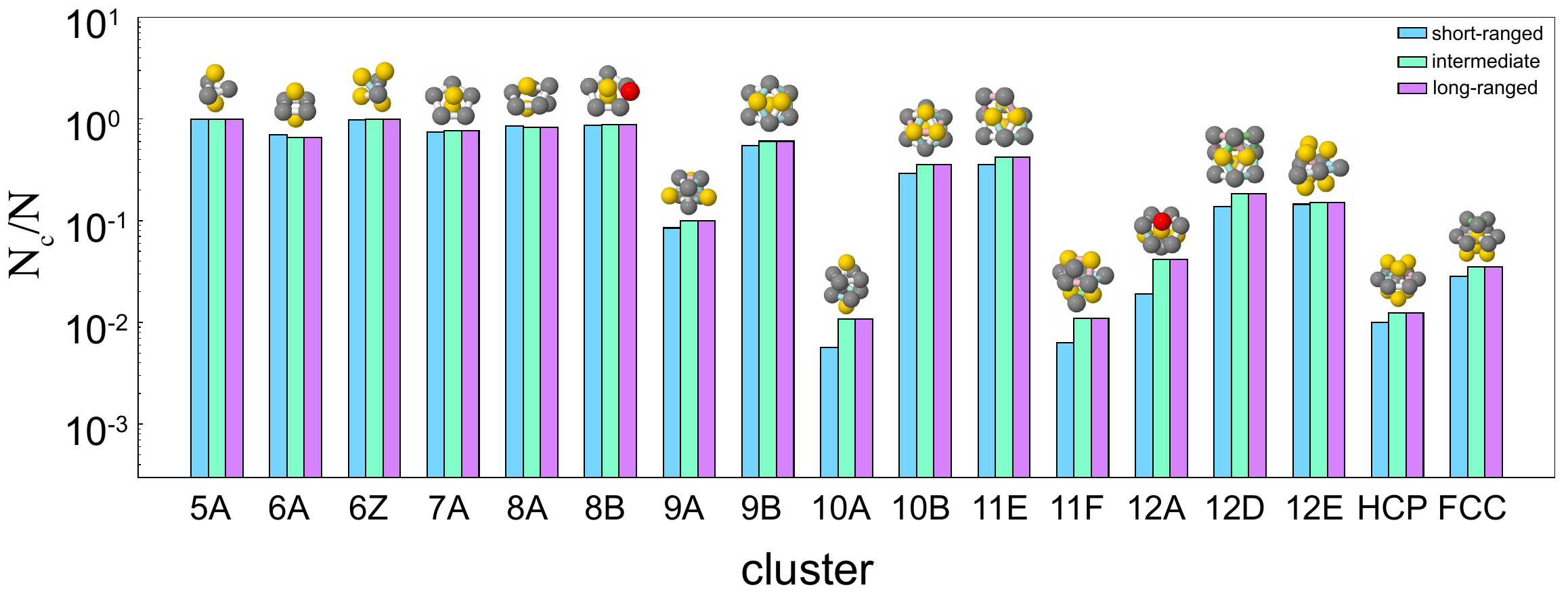}
\caption{Population of particles in clusters identified by the topological cluster classification for supersaturated fluids prior to nucleation. Shown are data  for the long-ranged system (violet) and short-ranged (blue). 
The state points shown are $\phi=0.3597$ and $0.2793$ for the long-ranged system and short-ranged systems respectively.}
\label{figTCCXtallising}
\end{figure*}

\begin{figure*}
\includegraphics[width=180 mm]{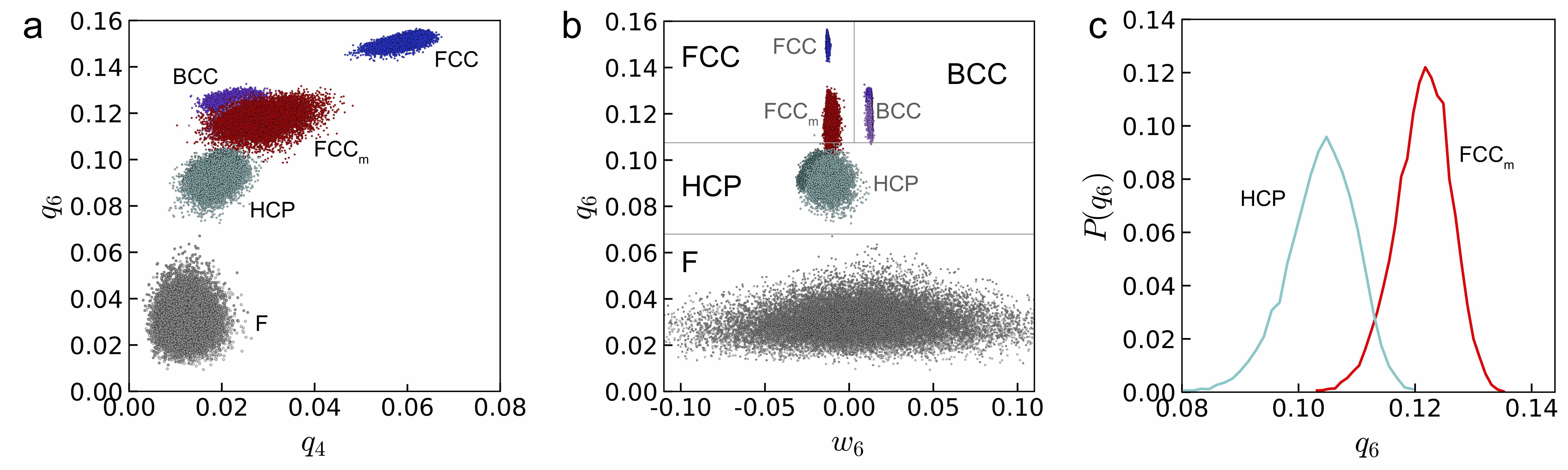}
\caption{Bond orientational order parameters for polymorphs considered.
(a) $q_4-q_6$ representation. 
(b) $w_6-q_6$ representation. 
In (a) and (b), the data are coloured as fluid, grey, BCC, violet, HCP, teal and FCC, blue. FCC at melting (FCC$_\mathrm{m}$) is shown in red.
Lines in (b) indicate regions identified with the structures indicated in black type. Grey type refers to data from simulations for each state point.
(c) Probability distribution of $q_6$ for the HCP (teal) and FCC at melting (red) states.
Here the teal line is HCP and the red is FCC at melting. Further details of the state points samples are given in Table~\ref{tableBop} in the Appendix.
}
\label{figBop}
\end{figure*}

\begin{figure}[h]
\includegraphics[width=75 mm]{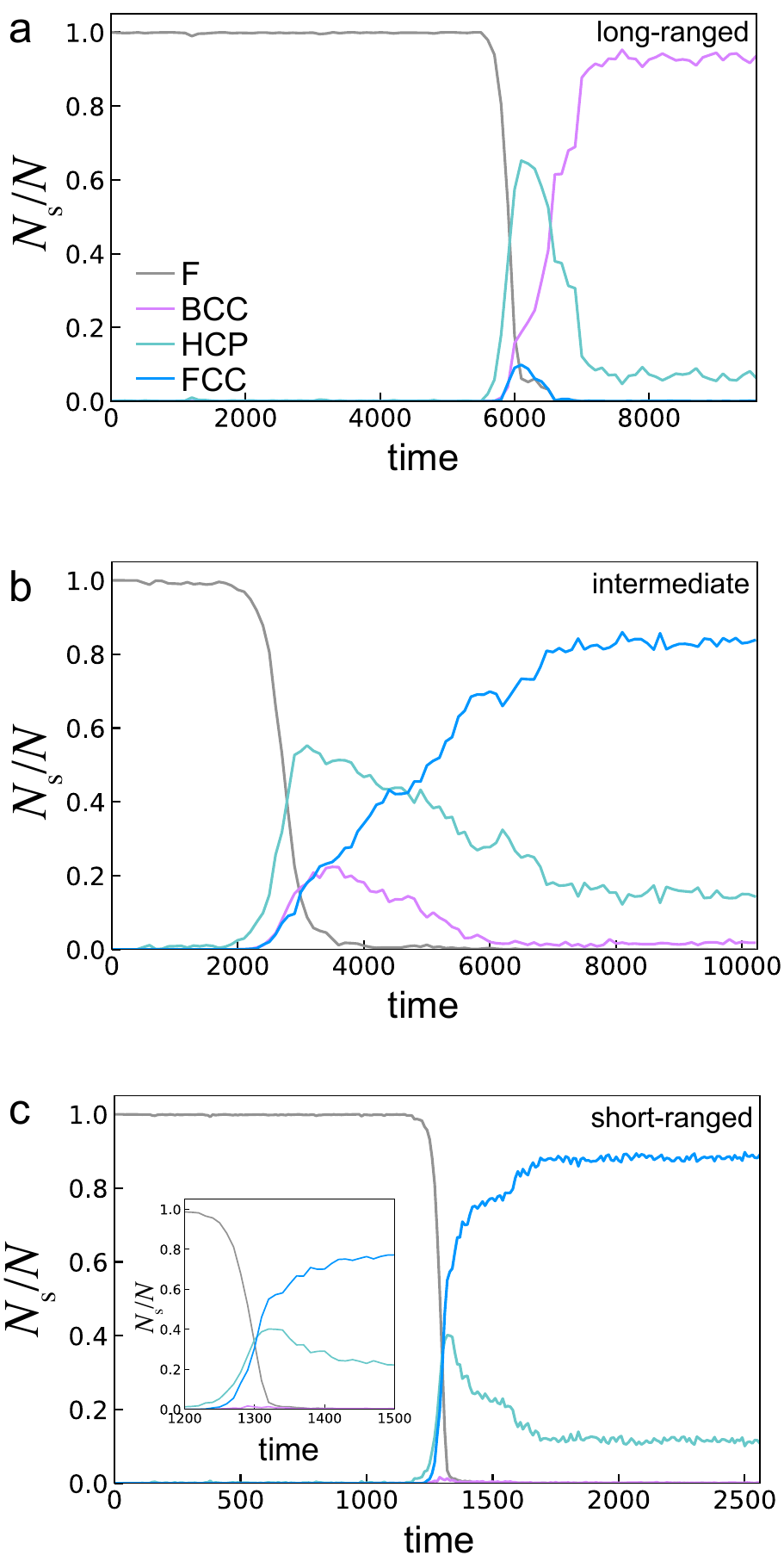}
\caption{
Time dependence of population of particles in different environments as characterised by the bond orientational order parameter.
(a) Long-ranged case. Here the initial volume fraction is $\phi=0.363$ and the pressure is $p=93.66 k_BT/\sigma^3$.
(b) Intermediate-ranged case. Here the initial volume fraction is $\phi=0.363$ and the pressure is $p=15.93 k_BT/\sigma^3$.
(c) Short-ranged case. Here the initial volume fraction is $\phi=0.279$ and the pressure is $p=8.368 k_BT/\sigma^3$.
Inset shows the same data zoomed in around the nucleation time.
Time is expressed in Lennard-Jones time units and the colours correspond to fluid, grey, BCC, violet, HCP teal and FCC, blue.
}
\label{figBopTime}
\end{figure}

For a complete explanation of the topological cluster classification~\citet{malins2013tcc} should be referred to. The topological cluster classification identifies target clusters by their bond network, these are polyhedra which are associated with a unique bond topology that corresponds to minimum (free) energy clusters of a given interaction potential. The bond network here is defined with a Voronoi decomposition combined with a distance criterion. As the basic building block for clusters, the algorithm constructs all the three, four and five-membered rings which can be constructed along the bond network.

In Fig.~\ref{figTCCStructures}, we show minimum energy clusters for the Morse potential. Here the grey particles indicate rings and the yellow particles are spindles Fig.~\ref{figTCCStructures} and the red are neither spindle nor ring particles. As shown in Eq.~\ref{eqMorse} in the Appendix, the Morse potential has a variable range, in that for a given number of particles, the topology of the minimum energy cluster may change for different values of the range parameter $\rho_0$. Here we follow the nomenclature of Doye \emph{et al}\cite{doye1995} where clusters corresponding to small values of $\rho_0$ take letters towards the start of the alphabet (eg 11A) with progressively larger values of $\rho_0$ given letters B.., ie 11B, 11C, etc. The 6Z cluster is the minimum energy cluster of the Dzugutov potential~\cite{malins2013tcc}. For consistency with the BOOP, we set the maximum bond length to the second nearest neighbours, although we note that for these dense systems, the Voronoi construction used will typically dominate in the determination of the bond network~\cite{malins2013tcc}. We set the parameter $f_c=0.82$ which controls the choice between three- and four-membered rings~\cite{malins2013tcc}.  In our analysis of the bulk liquids (Figs.~\ref{figTCCFreezing} and ~\ref{figTCCXtallising}), we have sampled at least $3\times10^6$ particles for each state point and so the statistical errors are relatively small with at least $3\times10^3$ sampled even for relatively rare clusters.

\subsection{Bond Orientational Order Parameters}
\label{sectionBond}

We have found the bond orientational order parameter (BOOP) analysis to be a reliable measure of particles in BCC environments~\cite{ingebrigtsen2019} and proceed to use the same method here. We also use the BOOP to identify particles in FCC and HCP environments, although very similar results are obtained using the TCC.

Here we have followed the method of  Lechner and Dellago~\cite{lechner2008} who obtained a clear distinction between the crystal structures of interest here by including second nearest neighbours.

\begin{equation}
Q_{lm} \equiv  \frac{1}{n_b}   \sum_\mathrm{bonds} Y_{lm}(\mathbf{r})
\label{eqQlm}
\end{equation}

\noindent
where $Y_{lm}(\mathbf{r})$ are the spherical harnonics with polar and azimuthal angles of the bond between a particle and its neighbours with respect to a fixed reference frame.

The invariant (with respect to the reference frame)

\begin{equation}
Q_{l} \equiv \sqrt{ \frac{4 \pi}{2l +1}  \sum_\mathrm{m=-l}^{l} |Q_{lm}(\mathbf{r})|^2  }.
\label{eqQl}
\end{equation}

The third-order invariants

\begin{equation}
W'_{l}  \equiv  \sum_{m_1,m_2,m_3}  \begin{Bmatrix} l  & l & l \\ m_1 &  m_2 & m_3 \end{Bmatrix}  Q_{lm_1} Q_{lm_2} Q_{lm_3} 
\label{eqWlprime}
\end{equation}

\noindent
and are normalised as 
\begin{equation}
W_{l}  \equiv \frac{W'_l}{ \sum_\mathrm{m} |\bar{Q}_{lm}(\mathbf{r})|^{3/2}  }.
\label{eqWl}
\end{equation}

Here,  in order to include contributions from the second nearest neighbours~\cite{lechner2008}, we set the bond length to be the second minimum of the radial distribution function $g(r)$ in the (supersaturated) fluid. 
In the case of crystal phases we take the fluid at coexistence, or, if the crystal is somewhat more compressed, we take the  second minimum of the $g(r)$ at coexistence and scale the range according to $(\phi/\phi_\mathrm{coex})^{1/3}$ where $\phi$ is the volume fraction of interest and $\phi_\mathrm{coex}$ is that at coexistence. To evaluate the BOOP, we use the BOP code of Wang \emph{et al}~\cite{wang2005}.

\section{Results}
\label{sectionResults}

\subsection{Higher-Order Fluid Structure}
\label{sectionHigher}

We begin our presentation of the results by considering the supercooled liquid prior to crystallisation. We compare the higher-order structure (as elucidated with the TCC) at freezing for 
the long-ranged (BCC stable) and short- and intermediate-ranged (FCC stable) cases in Fig.~\ref{figTCCFreezing}. Data for the long-ranged system (inverse Debye screening length ($\kappa\sigma=2$) are shown in violet, intermediate ($\kappa\sigma=4$) in green and the short ranged ($\kappa\sigma=10$) in blue. We see that there is very much more higher-order structure for the short-ranged system than the long-ranged case, with the intermediate-ranged case lying in between. For comparison, we also show a Weeks-Chandler-Anderson (WCA) system at freezing (which we take as an effective hard sphere volume fraction $\phi=0.4917$~\cite{royall2024}) for the same value of the WCA interaction strength $\beta \varepsilon_\mathrm{wca}=10$. This in fact has somewhat less higher-order structure than the short-ranged system, but rather more than the long-ranged case. We note that similar behaviour, of a ``lack of structure'' in a long-ranged system has been seen previously albeit for a different interaction potential and under a different mapping~\cite{taffs2010}.

In our simulations, we find that spontaneous nucleation occurs at rather different degrees of compression with respect to the phase boundary, as shown in Fig.~\ref{figPhase}. That is $\phi_\mathrm{nucl}/\phi_f=1.73$ for the long-ranged case while for the short-ranged system it is just $\phi_\mathrm{nucl}/\phi_f=1.18$, and $\phi_\mathrm{nucl}/\phi_f=1.29$ for the intermediate-ranged system where $\phi_f$ is the volume fraction of the fluid at freezing~\cite{hynninen2003}. This corresponds to a pressure scaled by that at freezing of $p/p_f=3.48$, $p/p_f=1.63$ and $p/p_f=1.67$  for the long-, short- and intermediate-ranged cases respectively.

Such a large supersaturation of the long-range system before spontaneous nucleation is observed may seem surprising, but it is compatible with the kinetic phase diagram of the Yukawa model~\cite{gispen2022}. In colloidal systems, the population of TCC clusters typically increases increases with volume fraction~\cite{taffs2013,royall2018jcp}. Here we find the same, in Fig.~\ref{figTCCXtallising}, where we show the populations of TCC clusters for simulations at the lowest volume fractions where the system crystallised for the three systems. 
We sample only the metastable fluid and to be confident that these are not influenced by nucleation, we take data only for $t<t_\mathrm{nucl}/2$ where $t_\mathrm{nucl}$ is the time at which the nucleus is first detected. Here population of TCC clusters is almost indistinguishable in all cases. This shows that the higher-order structure of the fluid before crystallisation is almost identical in all three cases.

\subsection{BOOP Analysis of Polymorph Selection}
\label{sectionBOOP}

\begin{figure*}
\includegraphics[width=175 mm]{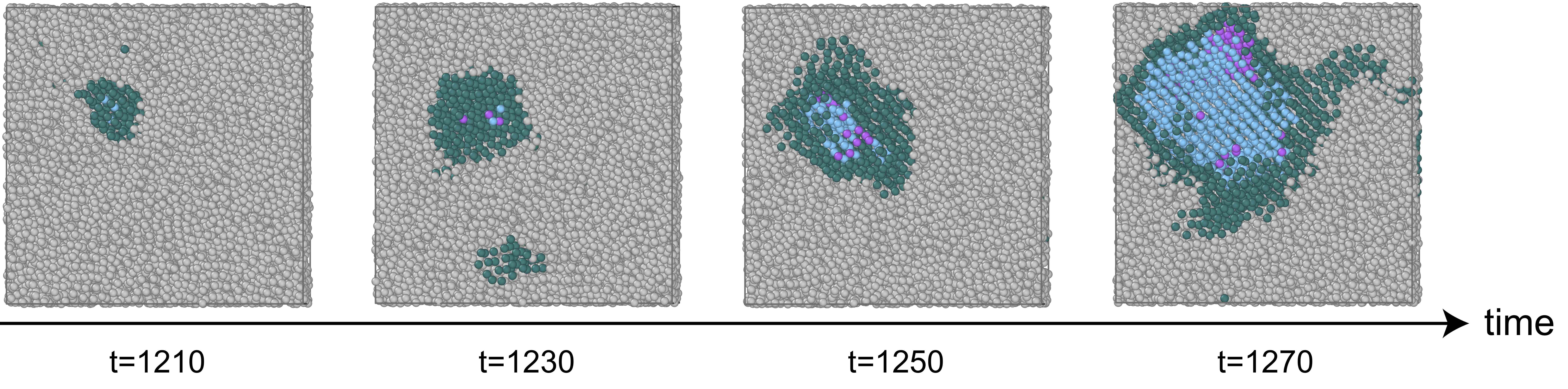}
\caption{Snapshots at selected times during a nucleation run showing (a) initial nucleus identified as HCP, (b) growing nucleus with small amounts of FCC and (c) emergence of BCC and FCC domains as the nucleus continues to grow. At (slightly) longer times an FCC domain decorated by HCP precursor particles is seen. Here particles identified as fluid are shown in grey, HCP in teal and FCC in blue. We consider the short-ranged system at an initial volume fraction of $\phi=0.279$. }
\label{figBopPretty}
\end{figure*}

We now turn our attention to the analysis with the bond orientational order parameters (BOOP). We begin by discussing the parameterisation of the BOOP, as this exhibited a somewhat unexpected behaviour, as shown in Fig.~\ref{figBop}. We implement the analysis of Lechner and Dellago (Sec.~\ref{sectionBond}). In particular, we consider the $(Q_4,Q_6)$ and $(W_6,Q_6)$ distributions.

To set parameters to distinguish the structure of interest, we performed NVT simulation of the relevant stable phases. We consider stable crystals of FCC, BCC and HCP and the fluid at freezing. For the HCP, we used the WCA potential at the (effective) volume fractions of 0.66 and 0.74. For our parameters, we found that this metastable crystal melted when the effective volume fraction was reduced significantly lower then 0.66. For the crystals, we used a ``magic number'' of particles to ensure a perfect crystal, 10976, 8192 and 8000 for FCC, BCC and HCP respectively. We used 10976 particles for the fluid and we considered two state points corresponding to the long ranged and short ranged case. For the BCC and FCC, we considered the melting volume fraction and also a somewhat higher volume fraction (around 25\% higher). A complete list of state points sampled is provided in Table.~\ref{tableBop} in the Appendix.

In the scatter plots shown in Fig.~\ref{figBop} (a and b), we see a clear distinction between the different crystal phases in both plots, and indeed the fluid. For the fluid at freezing, we saw no difference between the long-ranged, short-ranged or WCA systems and do not distinguish the points in Fig.~\ref{figBop}. Likewise the BCC and melting and at higher volume fraction showed no difference and are not distinguished. However in the case of the FCC at melting, although we saw no sign of melting in the simulations, the $(Q_4,Q_6)$ scatter plot in Figs~\ref{figBop}(a) shows that this state point is quite different to that at higher volume fraction. Indeed, the distribution of data points overlaps with BCC. This overlap disappears in the  $(W_6,Q_6)$ scatter plots and we use this to distinguish the polymorphs, as shown by the grey lines in Fig.~\ref{figBop}(b). No meaningful difference in the results was found in the case that we used the $(Q_4,Q_6)$ distributions to distinguish the crystal structures. The $Q_6$ distributions of the FCC at melting and the HCP are shown in Fig.~\ref{figBop}(c), showing that with our cut-off criterion of   $Q_6=0.107$ that there is some small overlap for the tail of each distribution. It is interesting to enquire as to the origins of the difference in the BOOP distributions in the FCC at melting and at higher volume fraction. While we leave a full investigation for the future, we did probe the structure with the TCC. In the case of the denser crystal, all particles were identified as FCC. In the case of the FCC at melting, 99.66\% of particles were identified as FCC so we conclude that these criteria of $Q_6,W_6$ are reasonable for our purposes.

We now analyse the process of nucleation using the BOOP. In Fig.~\ref{figBopTime}, we show the population of particles classified as supercooled liquid (grey), BCC (violet), HCP (teal) and FCC (blue). The results for the long-, intermediate- and short-ranged systems are shown in Fig.~\ref{figBopTime}(a), (b) and (c) respectively. In all cases, we see an increase in the HCP population just prior to the nucleation event when the system transforms to BCC (long ranged) or FCC (short and intermediate ranged). In the case of the intermediate-ranged system, we see a significant quantity of BCC, before this population falls as the FCC population continues to rise and the nuclei grow. We note that this state point lies close to the BCC-stable region in the phase diagram (Fig.~\ref{figPhase}). Similar behaviour is seen for other state points, as shown in Fig.~\ref{sFigBopTime} in the Appendix.

We therefore interpret these HCP regions as precursor nuclei to the stable phase which forms a little later. This is qualitatively similar to the results of \citet{lechner2011} and \citet{russo2012sm} who also found HCP precursors to  nucleation in the Gaussian core model for these same polymorphs.

A sequence of snapshots for the short-ranged system is shown in Fig.~\ref{figBopPretty}. Here we see that the HCP (teal) forms first, then initially small regions of mainly FCC (blue) with some BCC (violet) form within the nucleus.  At slightly longer times, more FCC particles are seen in the centre of the nucleus. As the nucleus grows, the centre becomes more dominated by FCC, surrounding by an HCP layer with a trace quantity of BCC at the FCC-HCP interface.

\begin{figure}
\includegraphics[width=85 mm]{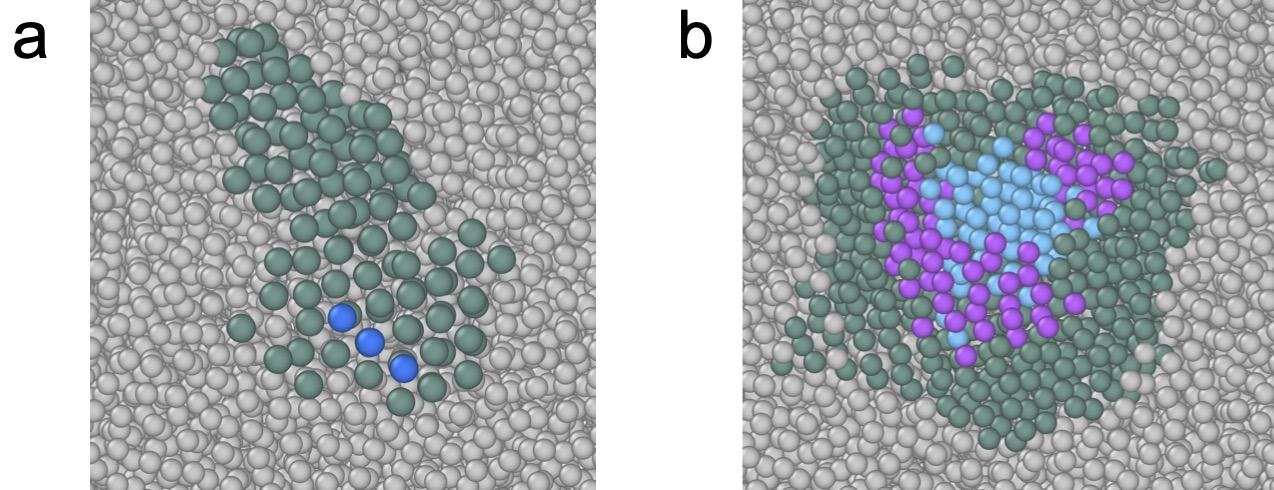}
\caption{Precursor particles with hexagonal order.
Particles identified as fluid are shown in grey, HCP in teal,  FCC in blue and BCC in violet.
(a) Long-ranged system at an initial volume fraction of $\phi=0.363$. This snapshot is taken at $t=5700$ prior to the formation of the BCC.
(b) Intermediate-ranged system at an initial volume fraction of $\phi=0.230$. This snapshot is taken at $t=2300$.
}
\label{figPrettyHexPrecursor}
\end{figure}

In Fig.~\ref{figPrettyHexPrecursor}, we show a snapshot of a precursor nucleus for the long-ranged (a) and intermediate (b) systems. The long-ranged case is for $t=5700$, and the time-evolution for this run is shown in Fig.~\ref{figBopTime}(a). We see that at this time, there is very little BCC, and that the ordering (interpreted with the BOOP) is HCP, ie precursors. In the intermediate-ranged case (b), the snapshot is taken at $t=2330$, by which time considerable amounts of BCC and FCC have formed (see Fig.~\ref{figBopTime}(b)), and indeed both can be seen in the middle of the ordered region. The coexistence of both polymorphs is broadly consistent with the experimental observation of a low surface tension between the two~\cite{chaudhuri2017}.

\begin{figure}
\includegraphics[width=90 mm]{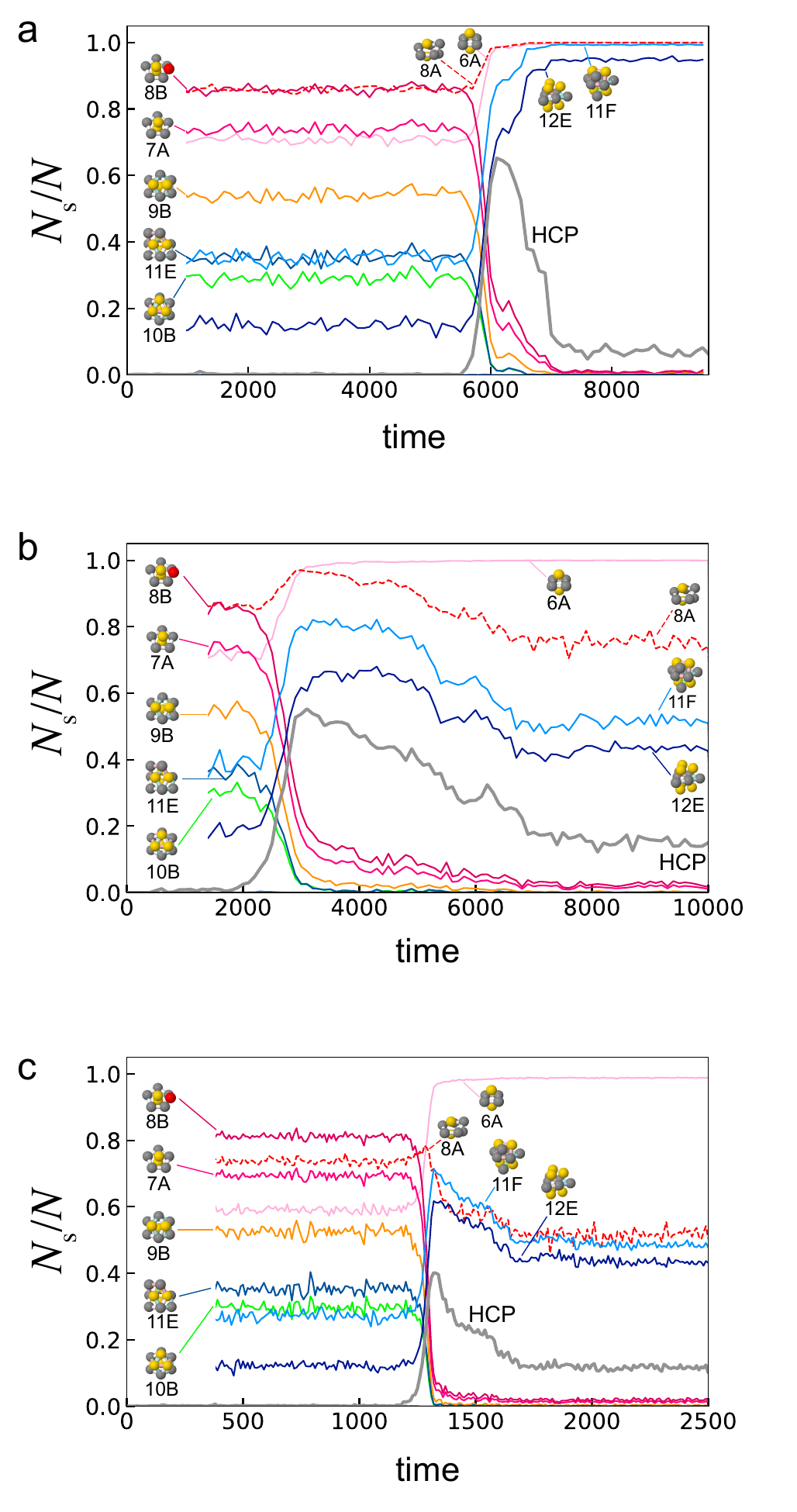}
\caption{Time dependence of population of particles in TCC clusters, Also shown is the population of HCP precursors identified with the BOOP.
(a) Long-ranged case. Here the initial volume fraction is $\phi=0.363$ and the pressure is $p=93.66 k_BT/\sigma^3$.
(b) Intermediate-ranged case. Here the initial volume fraction is $\phi=0.230$ and the pressure is $p=15.94 k_BT/\sigma^3$.
(c) Short-ranged case. Here the initial volume fraction is $\phi=0.279$ and the pressure is $p=8.368 k_BT/\sigma^3$.
}
\label{figTCCTime}
\end{figure}

\begin{figure}
\includegraphics[width=65 mm]{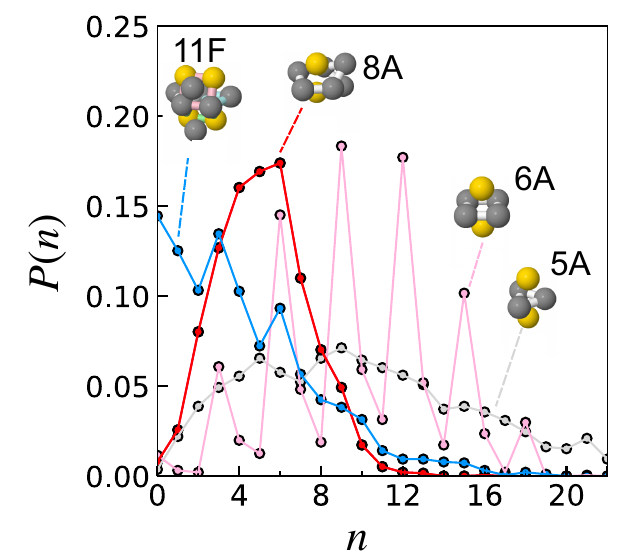}
\caption{TCC analysis of particles in HCP precursors. Shown is the distribution of the number of TCC clusters $n$ a precursor particle is found in.Here we show data for the long-ranged system.}
\label{figTCCDistrib}
\end{figure}

\subsection{TCC Analysis of Precursor Nuclei}
\label{sectionTCCprecursor}

What is the nature of these precursors identified as HCP? In Fig.~\ref{figTCCTime} we see in the time-evolution of the TCC cluster populations, that the populations of the 6A octahedron, 11F and 12E all increase just as the precursor forms. This occurs in both the longer-ranged [Fig.~\ref{figTCCTime}(a)], intermediate (b)
and  short-ranged (c) systems. In the long-ranged case, [Fig.~\ref{figTCCTime}(a)], more than half the system is identified in the HCP precursor state, and the populations of 6A,  8A, 11F and 12E all remain high. In the short-ranged run shown, the populations of these clusters was rather less at long times (presumably they are less compatible with the FCC here).

None of these clusters are strictly compatible with the HCP structure. We infer that the precursors, while they are hexagonally ordered (Fig.~\ref{figPrettyHexPrecursor}), may not by perfect HCP crystals. It is also worth noting that the TCC identification can tolerate some degree of distortion in the bond network. Inspection of the bond orientational order parameter analysis (Fig.~\ref{figBop}) shows that the HCP is found between the BCC (or FCC at melting) and the fluid, for both the ($Q_4,Q_6$) and ($W_6,Q_6$) representations. Therefore while the precursor undoubtedly shows considerable ordering, it may be distinct from being a full crystal phase.

We further probe the structure of the HCP precursors as follows. In Fig.~\ref{figTCCDistrib}, we show the distribution in the number of selected clusters a precursor particle is found in. (For the HCP cluster itself, particles are found in 1 or 2 HCP clusters.) We see a number of particles in up to 18 6A octahedra. Although the 6A is not strictly compatible with the HCP as noted above, some distortion is tolerated in the bond network. For larger clusters, the 8A has in fact some compatibility with the HCP crystal (though less than for the FCC~\cite{gispen2023acsnano}) and the 11F has a degree of HCP-like character~\cite{malins2013tcc}.
We note that the 5A triangular bipyramid is compatible with the HCP crystal. In Fig.~\ref{sFigTCCDistrib_150D}(a) in the Appendix we show the same analysis for the intermediate-ranged system, which shows a very similar distribution. The same holds for the short-ranged system (data not shown). We further show in Fig.~\ref{sFigTCCDistrib_150D}(b) the same analysis for a bulk HCP crystal. The distributions of the number of particles in the clusters is broadly similar, although there are some quantitative differences. However, at the level of this analysis it is hard to be sure whether these are true differences between the precursors and the bulk HCP or whether this relates to the fact that many of the precursor particles are at interfaces with the fluid, which of course is not the case for the bulk HCP (with periodic boundary conditions). This suggests that there is no significant difference in the structure of the precursors insofar that we can discern here. While further work is needed, we conclude that the precursors identified as HCP are consistent with an HCP crystal.

\begin{figure*}
\includegraphics[width=180 mm]{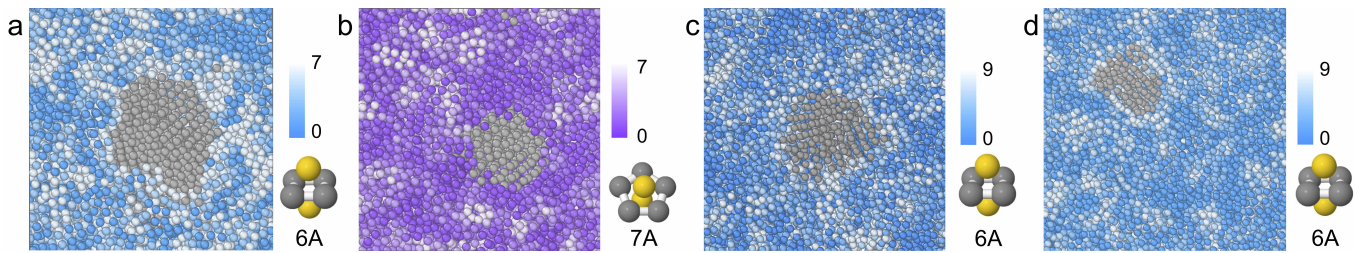}
\caption{Higher-order fluid structure around the hexagonally ordered precursors.
Here we show how many clusters of a specific geometry a particle is in. For example, in (a) a white particle is in 7 6A octahedra. 
(a) Short-ranged system, $\phi=0.279$ at $t=1100$. Grey particles are precursors. Blue shading denotes the number of 6A octahedra a particle is in.
(b) Long-ranged system, $\phi=0.363$ at $t=5700$. Grey particles are precursors. Violet shading denotes the number of 7A pentagonal bipyramids a particle is in.
(c) Long-ranged system, $\phi=0.363$ at $t=5700$. Grey particles are precursors. Blue shading denotes the number of 6A octahedra a particle is in.
(d) Intermediate-ranged system, $\phi=0.230$ at $t=2100$. Grey particles are precursors. Blue shading denotes the number of 6A octahedra a particle is in.
}
\label{figTCCNucl}
\end{figure*}

We now consider the effect of these precursor nuclei on the surrounding fluid. It has previously been shown that FCC nuclei tend to suppress local fivefold symmetric order in the liquid for hard spheres~\cite{gispen2023acsnano}. Indeed the formation of FCC may be controlled by adjusting the degree of fivefold symmetry in the liquid~\cite{taffs2016}. We perform the same analysis as that carried out by ~\citet{gispen2023acsnano}, to show the effect of the hexagonally ordered precursor nucleus on the liquid. Here, then we find that, like the case for hard spheres, there is significant suppression of fivefold symmetry in the liquid due to the hexagonally ordered nucleus [Fig.~\ref{figTCCNucl}(b)]. That is, the population of 7A pentagonal bipyramids are suppressed close to the nucleus as indicated by the darker violet.

It was also shown that there was an enhancement of the dodecahedral 8A cluster on the edge of FCC nuclei, which was argued to promote the selection of that polymorph. Here for these HCP-like precursors, instead we see an enhancement of 6A octahedra relative to the fluid [Fig.~\ref{figTCCNucl}(a,c,d)]. (The same rendering for the 8A dodecahedron is shown in Fig.~\ref{sFig8ANucl} in the appendix).

\section{Discussion}
\label{sectionDiscussion}

Like Russo and Tanaka~\cite{russo2012scirep} with hard spheres and the Gaussian Core Model~\cite{russo2012sm}, and Lecher and Dellago with the GCM 
~\cite{lechner2011}, we find a hexagonally ordered precursor to the formation of the BCC and FCC nuclei in the 
Yukawa system with a slightly softened core. Our results are also compatible with the experimental work of Tan \emph{et al.}~\cite{tan2014}. They are not in alignment with work that found no precursors in the hard core Yukawa system~\cite{dejager2023}, although that work used a rather somewhat different approach. It seems that a systematic study of different model systems with a range of order parameters would be desirable in the near future.
Nevertheless some comments regarding the apparent discrepancy between this work and that of de Jager \emph{et al.}~\cite{dejager2023} are in order. That work considered somewhat different parameters, namely a fully hard core and a contact potential of $\beta\varepsilon_\mathrm{yuk}=81.0$. Perhaps more significantly, a different implementation of the BOOP was employed, namely the so-called solid angle nearest neighbour method, rather than the next-nearest neighbours considered here. While a proper investigation would be needed to be sure, It is possible that the precursors identified here as HCP were identified as the stable crystal in that work.

The regime of supersaturation that we have explored is relatively small (Sec.~\ref{sectionMolecular}). A key challenge then is to determine over what domain of supersaturation that such precursors are found. Considering the results of Mithen \emph{et al.}~\cite{mithen2015} for the GCM, then it is possible that the hexagonally ordered precursors might not be found at weaker supercooling.

In the future, it would be attractive to perform simulations at weaker supersaturations than has been possible here, using for example forward flux sampling or umbrella sampling. This would enable one to probe to what extent these hexagonally ordered precursors are found as the degree of supersaturation falls. Given the drop in higher-order structure in the BCC-stable long-ranged case, notably the disappearance of hexagonally ordered HCP and FCC (Fig.~\ref{figTCCFreezing}, violet data), it is tempting to imagine that the proposal of Alexander and McTague of BCC ordering in general~\cite{alexander1978} may be expected, consistent with the findings of refs~\cite{desgranges2007jcp,krazter2015}.

We have found that the TCC cluster populations, as a measure of the higher-order structure are very similar for all three systems for state points where nucleation was found on the simulation timescale. The TCC clusters are themselves minimum energy clusters of the variable-ranged Morse potential, Eq.~\ref{eqMorse}, Fig.~\ref{figTCCStructures}. Now of course the interactions here are not Morse interactions and indeed they are even repulsive. Yet it is possible, in the spirit of the WCA treatment of the Lennard-Jones interactions, ie to truncate and shift at the minimum of the potential, to do the same with the Morse interaction (Eq.~\ref{eqTMR} in the Appendix). It turns out that such truncated Morse interactions have very similar higher-order structure to the full potential, as indeed the WCA has to the full Lennard-Jones potential~\cite{taffs2010}.

Such a ``truncated Morse'' interaction may then be compared with the Yukawa interaction used. We find values of the Morse range parameter $\rho_0 \approx 2.5$ for the longer-ranged, $\kappa\sigma=2.0$ case and $\rho_0 \approx 4.0$ for $\kappa\sigma=10.0$. For reference, the Lennard-Jones interaction corresponds to a still shorter interaction of  $\rho_0 \approx 6.0$~\cite{doye1995}. Now it turns out that the minimum energy structures for  $\rho_0=2.5$ and $\rho_0=4.0$ are identical except for $m=11$ and $m=12$. So even applying this line of reasoning to the populations of TCC clusters, we would in fact expect them to be similar here, despite the change in the interaction range. This is what we find. Therefore, it seems reasonable that polymorph selection here occurs in the hexagonal precursor state. Determining why this hexagonally ordered state can form both polymorphs stands as an interesting question for the future.

It is interesting to consider the transformation of the HCP precursors into BCC and FCC. Similar solid-solid transformations have been considered previously~\cite{li2021}. The extent to which these are thermally activated would be an interesting question. In principle it is not expected that HCP forms in the first place. It this soft matter system we speculate that its transformation to the stable polymorph is thermally driven, and would be an intriguing  question to probe in detail in the future.

\section{Conclusion and Outlook}
\label{sectionConclusion}

We have studied nucleation in a model polymorphic system using molecular dynamics simulations. We find a hexagonally ordered precursor to both the BCC and FCC crystals, each of which forms (after the precursor) when it  is the stable phase. The BCC is stable for a long-ranged interaction (here the inverse screening length $\kappa \sigma =2.0$) while the FCC is stable for an intermediate ($\kappa \sigma = 4.0$) and  short-ranged interaction ($\kappa \sigma = 10.0$), Fig.~\ref{figPhase}. The precursors suppress the fivefold symmetry of the surrounding fluid in a similar manner to FCC nuclei in hard spheres~\cite{gispen2023acsnano}.

In our simulations which examine spontaneous nucleation (on timescales relevant to experiments with colloids~\cite{taffs2013,tan2014}), a much higher supersaturation is needed to observe nucleation for long-ranged interactions than is the case for shorter-range interactions. This is consistent with our observation of a fluid with much less order at freezing in the case of a long-ranged interaction when compared to a shorter-ranged case.

In the future, it would be attractive to carry out this kind of analysis on state points with weaker supersaturation than are accessible to our direct simulations. Determining any difference between the precursors in the case that FCC is nucleated and BCC is nucleated stands as an interesting challenge. While the precursors in both satisfy our criteria for HCP (Fig.~\ref{figBopTime}), it is tempting to imagine that there is some subtle difference in their structure which somehow encodes that one forms FCC and the other BCC. However, our analysis found no such difference. In addition, other systems, for example colloids with attractive interactions could be investigated with the same methodology. It would also be interesting to extend this approach to other classes of materials. (Non-hexagonally ordered) precursors have been found in water~\cite{russo2016}, NiAl~\cite{hu2021} and may be relevant in more complex systems such as calcium carbonate solution, a model system for biomineralisation~\cite{demichelis2011}.

\begin{acknowledgments}
It is a pleasure to thank Marjolein Dijkstra for many inspiring conversations around nucleation, hard core Yukawa systems and polymorph selection, and Bob Evans, Daan Frenkel, Yi-Yeoun Kim, Fiona Meldrum, David Quigley, John Russo, Richard Sear Frank Smallenburg and Hajime Tanaka for many conversations on the topic of nucleation and polymorph selection. CPR acknowledges the Agence Nationale de Recherche for grant DiViNew.
\end{acknowledgments}

\vspace{20pt}

\section*{Appendix}
\setcounter{equation}{0}    
\renewcommand\theequation{A\arabic{equation}} 

The topological cluster classification identifies minimum energy clusters. Here we consider minimum energy clusters from the Morse potential, which reads

\begin{equation}
\beta u_\mathrm{morse}(r) =
\beta \varepsilon_\mathrm{morse} \left[ \mathrm{e}^{ - 2 \rho_0 (r-\sigma)} - 2 \mathrm{e}^{ -  \rho_0 (r-\sigma) }  \right]
\label{eqMorse}
\end{equation}

\noindent
where $\rho_0$ controls the range of the interaction, and, thus the topology of the clusters in Fig.~\ref{figTCCStructures} as determined by \citet{doye1995}.

It is possible, in the spirit of the WCA treatment of the Lennard-Jones interaction, to define a truncated and shifted Morse potential which retains only the repulsive part.

\begin{align}
\label{eqTMR}
u_\mathrm{tm}(r) &= 
\begin{cases}
\beta \varepsilon_\mathrm{morse}  \left[ \mathrm{e}^{ - 2 \rho_0 (r-\sigma)} - 2 \mathrm{e}^{ -  \rho_0 (r-\sigma) } +1 \right] &\ r \leq \sigma \\
0 &\ r > \sigma. \\
\end{cases}
\end{align}

One can compare $u_\mathrm{tm}(r)$ to the Yukawa potential with the softened core Eq. ~\ref{eqHYuk} and select a value of $\rho_0$ which best matches Eq. ~\ref{eqHYuk}.
We arrive at $\rho_0\approx 2.5$ for the long-ranged case $\kappa\sigma=2.0$ and $\rho_0\approx 4.0$ for the long-ranged case $\kappa\sigma=10.0$. For the full Morse potential, $\rho_0\approx 6.0$ is rather close to the Lennard-Jones interaction.

\begin{table}[]
\begin{tabular}{|l|c|c|l|}
\hline
State & $\phi$ & $\kappa \sigma$ \\ \hline
fluid & 0.2368 & 10.0  \\
fluid & 0.208 & 2.0  \\
fluid & 0.492  & WCA \\
FCC$_\mathrm{m}$ & 0.2468 & 10.0   \\
FCC & 0.35 & 10.0   \\
HCP & 0.661 &  WCA \\
HCP & 0.741 & WCA  \\
BCC & 0.21 & 2.0 \\
BCC & 0.315 & 2.0  \\ \hline
\end{tabular}
\caption{State points used to parameterize the BOOP in ~Fig.~\ref{figBop}. The lower volume fraction for BCC and FCC were taken at melting. }
\label{tableBop}
\end{table}

\begin{figure*}
\includegraphics[width=160 mm]{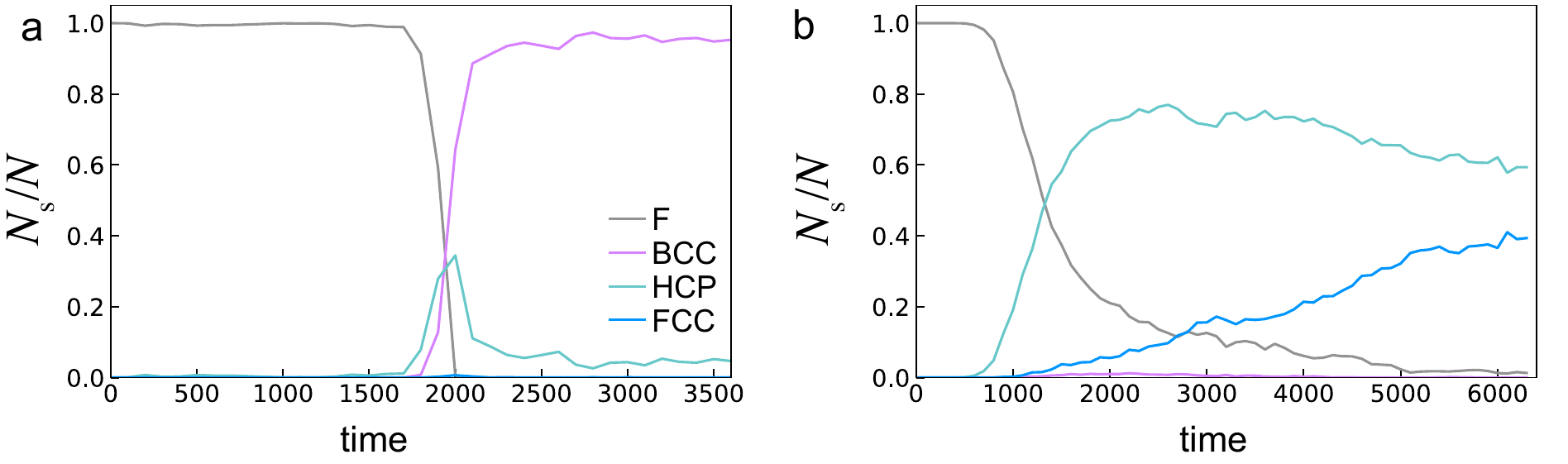}
\caption{
Time dependence of population of particles in different environments as characterised by the bond orientational order parameter.
(a) Long-ranged case. Here the initial volume fraction is $\phi=0.377$  and the pressure is $p=95.64 k_BT/\sigma^3$.
(b) Short-ranged case. Here the initial volume fraction is $\phi=0.2813$.  and the pressure is $p=9.564 k_BT/\sigma^3$.
Time is expressed in Lennard-Jones time units and the colours correspond to fluid, grey, BCC, violet, HCP teal and FCC, blue.
}
\label{sFigBopTime}
\end{figure*}

\begin{figure*}
\includegraphics[width=140 mm]{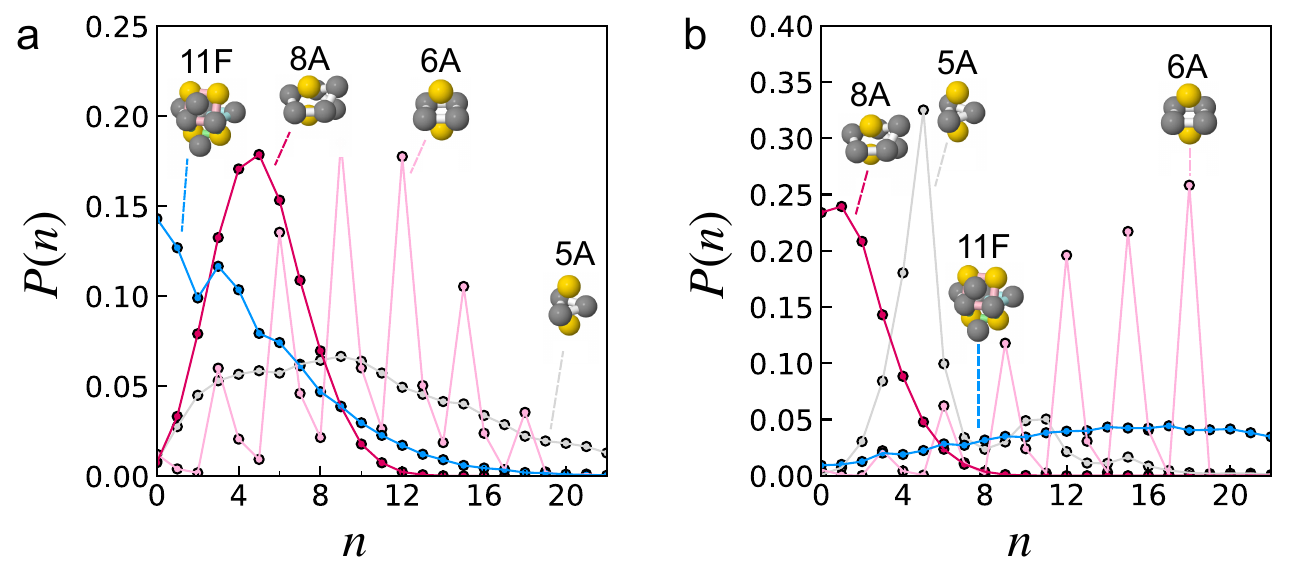}
\caption{TCC analysis of particles in HCP precursors and bulk HCP crystal. Shown is the distribution of the number of TCC clusters $n$ a precursor particle is found in.
Here we show data for the intermediate-ranged system (a) for $\phi=0.230$ and a bulk HCP crystal (b) where $\phi=0.661$.}
\label{sFigTCCDistrib_150D}
\end{figure*}

\begin{figure}
\includegraphics[width=55 mm]{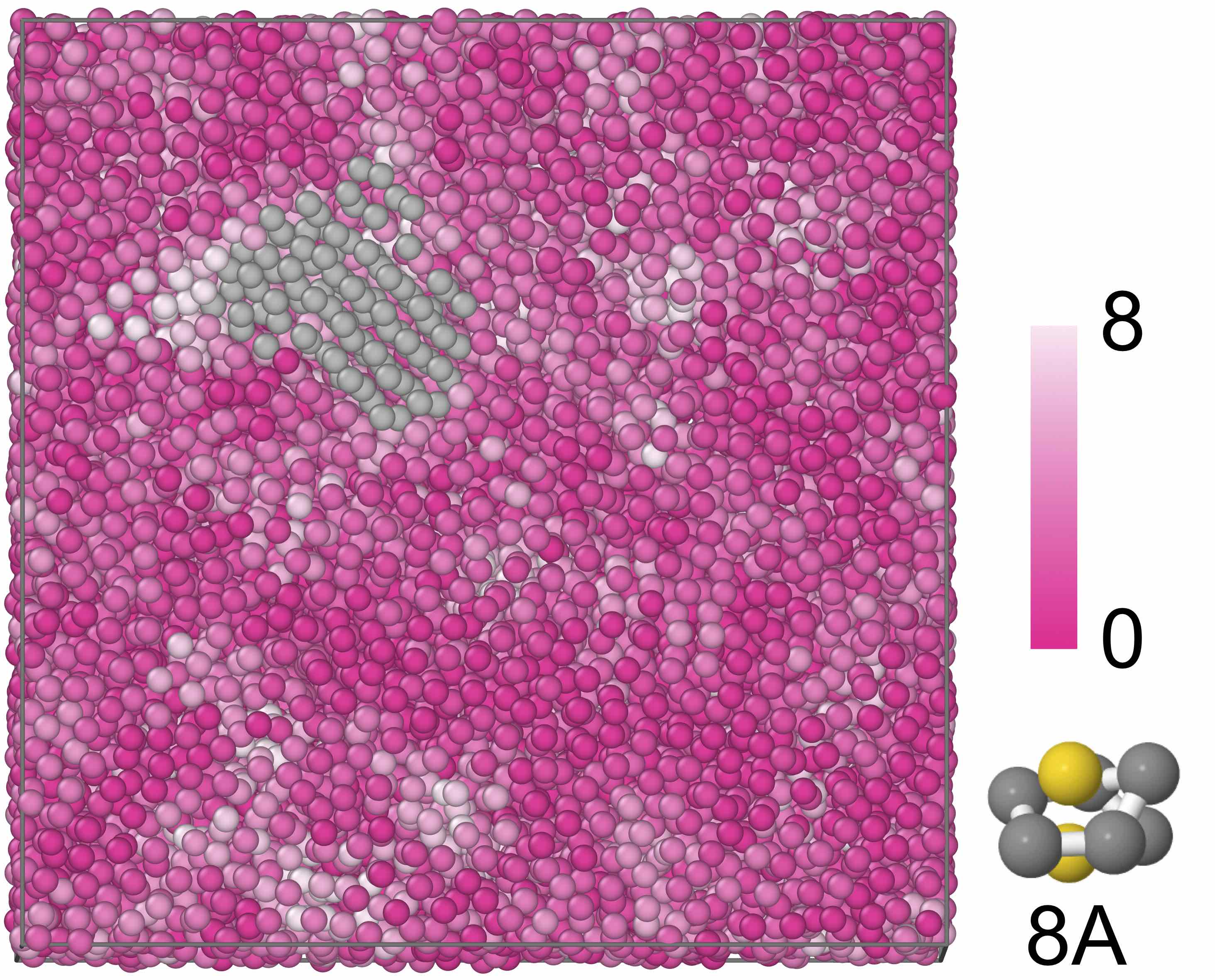}
\caption{Rendering showing how many 8A dodecahedra a particle is in. Short ranged system, $\phi=0.279$. 
Grey particles are precursors. Pink shading denotes the number of 8A a particle is in.}
\label{sFig8ANucl}
\end{figure}

\end{document}